# *A survey on Blockchain-based applications for reforming data protection, privacy and security*

Phan The Duy, Do Thi Thu Hien, Van-Hau Pham

Information Security Laboratory
University of Information Technology, VNU-HCM
Ho Chi Minh city, Viet Nam
e-mail: {duypt, hiendtt, haupv}@uit.edu.vn

*Abstract*— The modern society, economy and industry have been changed remarkably by many cutting-edge technologies over the last years, and many more are in development and early implementation that will in turn led even wider spread of adoptions and greater alteration. Blockchain technology along with other rising ones is expected to transform virtually every aspect of global business and individuals' lifestyle in some areas. It has been spreading with multi-sector applications from financial services to healthcare, supply chain, and cybersecurity emerging every passing day. Simultaneously, in the digital world, data protection and privacy are the most enormous issues which customers, companies and policy makers also take seriously into consideration due to the recent increase of security breaches and surveillance in reported incidents. In this case, blockchain has the capability and potential to revolutionize trust, security and privacy of individual data in the online world. Hence, the purpose of this paper is to study the actual cases of Blockchain applied in the reformation of privacy and security field by discussing about its impacts as well as the opportunities and challenges.

*Keywords: blockchain; privacy; data protection; security; blockchain adoption.*

## I. INTRODUCTION

In the past years, blockchain has become famous as the state-of-the-art technology behind cryptocurrencies such as Bitcoin or Ethereum which enables firms to record information or complete product history for both businesses and consumers throughout the entire progress on a distributed ledger, in a secure, irrevocable and immutable manner. Instead of storing data in a single location by uploading it to a central server, blockchain splits everything into small chunks and distributes them across the entire network of computers.

Due to the with huge market potential, blockchain and its underly technology have become a promising area to investigate. Many publications have been made on different aspects of Blockchain. In [1], authors focus on the application of Blockchain in a specific field – financial institutions in Korea in 2017. Another survey [2] makes an overview of security and privacy issues in Blockchain as well as their impact in different trends and applications.

Moreover, while we all reap the benefits of a data-driven society, there is an upward trend of public concern about user privacy. Since some privacy leakage scandals like the story about NSA's surveillance [3] or Facebook & Cambridge Analytica [4], privacy-preserving has been gained more awareness of everyone, not only customers but also corporations, especially. Data as the valuable asset, has been collected and used by companies to personalize services, predict future trends, optimize the corporate decision-making process and more. This proves that it plays an essential role in the economy, society and modern lives, such as critical infrastructure, medical equipment and even autonomous cars. However, not all of data is disclosable due to individual characteristics and sensitive level, it is also unlimited in a specific type or area which can be corporate financial data, personally identifiable information, daily individual activities and medical records. Since the global data volume is exploding exponentially with the growth of Big Data and metadata, data analyzing is more and more frequent, efficient, flexible with the supports of machine learning, natural language processing and artificial intelligence. For that reasons, applications with more restricted privacy or participation requirements strictly access to the network.

Adaptability of blockchain for reforming data protection, privacy and security can be considered as an important motivation to research and develop new technologies. However, it also needs to note that blockchain is not always the best solution all the time, in some cases it is more suitable to utilize traditional centralized databases. Hence, the purpose of this paper is not only to take an overview of blockchain – a rising technology that is experiencing explosion of interest but also investigate potential blockchain-based applications, as well as its actual implements and major challenges of applicability in security and privacy.

The structure of this survey is conducted in five sections. Some basic concepts of blockchain and its underlying techniques will be described in section 2. Section 3 focuses on the blockchain technology in the context of privacy and security to give an overview of the applicability in this field. Section 4 discusses future direction research for the applicability of blockchain. We make some conclusions in Section 5.

## II. BLOCKCHAIN CONCEPTS

### A. Blockchain overview

The centerpiece of Blockchain technology is decentralized structure, considered by advocates to be tamper-proof, immutable, time-stamped record keeping and more efficient. There are many techniques underlying this terminology consisting of cryptography, mathematics, networking and the sharing economy model. As the backbone of blockchain's operations, instead of centralized server, a peer-to-peer (P2P) model with consensus algorithms is used to take control whole of the network to address the problem of synchronization from distributed database. It is decentralized environment over the network and does not depends on any third-party to be maintained.

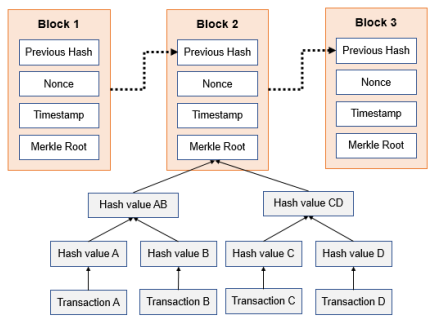

Figure 1. Linked list connection between blocks in blockchain

Overall, blockchain is considered as a distributed ledger used to store transaction records. The goal of Blockchain is to provide anonymity, privacy, security and transparency to all its users in chronological order of their verification, by removing a role of the central authority. The data that needs to be stored in blockchain, would be broadcasted into the whole P2P network to be verified by the certain consensus algorithms. After receiving the admission of all nodes, the new block containing objective data is added to the blockchain. All records of this ledger can be accessed by every participant in the blockchain network, but it is unchangeable upon the data has been approved by all nodes. Consequently, it is more difficult, even infeasible for hackers to modify block's information, since they must have controls over multiple systems to overcome consensus mechanism.

When it comes to the structure and components of blockchain, to ensure secure integrity and reliability of transaction records without third-parties, the cryptographically secure mechanisms including Hash, Digital Signature and Cryptography play the important roles. The simplified structure of blockchain with the most important components is showed in Figure 1. A Merkle tree, also known as a binary hash tree, is a data structure used for efficiently summarizing and verifying the integrity of large sets of data [5]. In addition to the timestamp and transaction data, each block contains a hash of the prior block, and a "Nonce" value, working as a digital fingerprint linking the blocks to form a blockchain. This chain of transactions cannot be tampered with, which the ledger is so secure.

### B. Types of Blockchain

Blockchain, at its first launch day with purpose of leveraging Bitcoin, is a permissionless technology. Since then, to meet the growing requirements of use cases, other types of blockchains have been created, permissioned blockchains. With permission control approach, the access rights to a blockchain will be limited to the parties involved in the creation of that particular network, or those granted access (read/write/verify) to it by the creating parties.

The entities participating in Blockchain operations consists of two groups of *readers* who can access to join the network and *writers* validating candidate blocks by consensus algorithm. According to assess and managing permission, blockchain can be typically categorized mainly three forms, including public blockchain, private blockchain and consortium blockchain, which emerging after Bitcoin introduced blockchain to the world [2] [6].

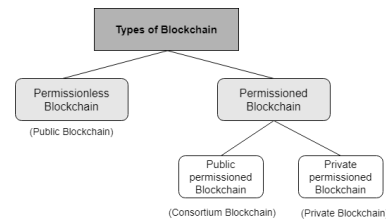

Figure 2. Different types of blockchain

Regarding to anonymity of validators, one blockchain structure can be classified into public or private manner. On the contrast, trustworthy of validators is measured by permissionless or permissioned model, which can be used to identify the type of specific blockchain. While public or permissionless blockchains are open, decentralized and slow, private/permissioned blockchains are more centralized and closed, either partially or completely.

*Public blockchain* has the greatest benefit of publicity, allowing any individual may access to be a part of the ecosystem to maintain distributed ledger. The entire public blockchain system is not only open and transparent, but also the identity of each node may remain anonymous. All participants of this kind can freely access information, submit and confirm transactions effectively. Also, the anonymity and privacy of the nodes in the system are therefore protected, like Bitcoin, Ethereum, Litecoin... This leads to the decrement of the performance in verification and consensus status.

On the contrary, *private blockchain* as its name suggests, provides a private property to an individual or organization where controlling read-write access belongs to only one organization. It is also called Permissioned Blockchain, for example, Hyperledger [7], Multichain [8], MONAX [9]. Access rights are restricted by granting permission to a group of participants. The sole distinction between public and private blockchain is related to who is allowed to participate in the network, execute the consensus protocol and maintain the shared ledger. Comparing to public kind, private one possesses important advantages of transaction cost reduction and easier-data handling. However, this type still faces with the single point of failure in the centralized management system.

Therefore, the *consortium blockchain* – the hybrid type of two prior categories – is partly decentralized. Take XDC Hybrid Blockchain [10] as an indication, it is consortium blockchain created by a Singaporean company XinFin. It has demonstrated itself up to leverage the power of both the public and private types. Moreover, there are other examples of Consortium Blockchain, including, EWF (energy) [11], R3 [12], Corda [13]. To be more precise, Consortium blockchain refers to a blockchain with consensus procedures controlled by pre-defined nodes, which means operating under the management of more than one organization which in a way withstands the whole network against a single point of failure. An association, typically of several participant such as banks, government entities, hospitals, and so on, can use consortium type to ensure the highest levels of privacy and security through a set of predetermined parties or identities. Hence, consortium blockchain can be used for such trades or activities between a federation of organizational partnerships to remove costly friction in business transactions with significant benefits.

These different types of Blockchain target a specific set of use cases for different domains of industries depending on their characteristics. Organizations or enterprises must consider the tradeoffs when they want to choose a relevant type of blockchain for their purpose. If the highest levels of privacy and control are required, Private and Consortium Blockchain are the best solutions. But once openness and anonymity are necessitated, Public Blockchain is more relevant.

*C. Social concerns on Blockchain*

Being a disruptive technology with many under-covered and unpredictable chances as well as issues, Blockchain still faces with the legal regulation's acceptance or policy normalization. This is the most concerned challenge for bringing blockchain into many domains. However, along with remarkable development and applicability in many sectors, human resource demand for blockchain is increasing significantly but in short supply. It is the most fastest growing skill with more than 6000 percent growth over a year [14], which is sought by not only new growing companies but also traditional ones desiring to catch up this technology.

To meet Blockchain workforce demands, lots of educational organizations, universities have already gotten into the race of this promising technology in their courses and academic program [15], such as The Australian University RMIT with a course named Developing Blockchain Strategy [16], "Blockchain: Foundations and Use Cases" course [17] of blockchain start-up ConsenSys Academy, Coursera [18], IBM with "IBM Blockchain Foundation for Developers" [19]. Besides that, governments also have their preparation to apply blockchain, an example is South Korean with their own 6-month course for training blockchain experts [20]. Due to the lack of trained experts, fresh computer science graduating from top engineering schools like UC Berkeley, Stanford, MIT CS or people attending a coding school are also potential workforce for many startups [21].

III. BLOCKCHAIN-BASED REFORMATION IN DATA PROTECTION, PRIVACY AND SECURITY

A handful of promising and potential blockchain-based applications have already been launched recent years. However, when considering blockchain as a solution in a specific application scenario, it is vital to keep in mind that it might not be the most appropriately technical approach. This part gives a brief view on realistic use cases of blockchain recently in data protection, privacy and security.

*A. Identity management*

Identity management systems (IdM systems) take responsibility for managing a collection of identities, their authentication and some related information [22], to assure the capability of enabling the right individuals to access the right resources at the right times and for the right reasons. It takes in mind some key terminologies as objects in its operation. An entity can be an individual, a device, a subject or an object, or anything with distinct existence, it can have a collection of multiple characteristics called identity, which distinguish it from other entities. For identification purpose, some of those identities will be used as credential. There are various types of credentials, the most used example is password or other secret information agreed between two sides. Entity identifier is a unique index used by a system to refer to associated identity [23].

For a traditional IdM system, some important requirements need to be met and considered in implementation, including privacy, security and usability. Most IdM systems are large, centralized database repositories located in large network or cloud servers [24], which can be experienced a single point of failure. Moreover, identification information can be considered as critical data, protection requirements need to be covered not only when transferring through the network but also in the storage mechanism. Although solutions from third-party can simplify some operations of the system, trust between IdM and functions provided by these organization is required. Despite that these problems have been somehow solved, there are still some critical aspects such as user privacy and anonymity of user's identity and data, which lead to the emergence of a new approach based on Blockchain technology.

A Blockchain IdM system based on public identities ledger invented by Sead Muftic has been granted a patent in US [24]. This invention proposed "an identity management system managing digital identities which are digital objects containing attributes used for the identification of persons or other entities in an IT system and for making identity claim". Based on the idea of Bitcoin public ledger and the principle of Blockchain technology, identity objects are maintained in a constructed global, distributed, append-only public identities ledger and linked to other objects in the chain using digital signatures. These objects are also encoded, cryptographically encapsulated before being added into the ledger. An identity object can be validated in permissioned or unpermissioned mode. Permissioned mode needs special entities called BIX Security Policy Providers to validate the binding of digital identities to the associated entities based on given policies. However, in unpermissioned mode, this component can be ignored, members can mutually validate their entities. The validation of identities and creation of personal identities chain as branches of public identities ledger reflect the relationship with validators of their identities. Identities in the ledger are followed by identities that they have validated, so that this system can have multiple identities chain linked to different BIX Security Policy Provider. An example is the case of business entity of an organization or a department can validate the identities of its user and binding to the real persons.

In [25], Ori Jacobovitz has made a discussion on Blockchain technology and focused on its application in identity management. The author overviews multiple identity-related solutions based on Blockchain and classifies them into different groups according to their importance/relevance. For example, Bitnation [26] - a governance platform powered by blockchain technology that intends to provide government services in a decentralized and voluntary manner without any geographic bounds, establish the concept of world-citizenship. Any individual can become a citizen of Bitnation by signing the constitution. Whereas that, e-Residency [27] launched in Estonia enables people from other places to access to services in Estonia such as company formation, banking, payment processing and taxation. The identity information used in this system can be vary from national passport, photographs, birth certificates, business contracts… Application of this system is targeted towards location-independent entrepreneurs. IBM Blockchain Trusted Identity [28] is another example for Blockchain application in identity management. This solution creates a decentralized approach for identity management building on top of open standards.

Furthermore, Dubai's Immigration and Visas Department has consented to arrangement with UK-based blockchain startup to test its "gate-less" border technology for passenger at Dubai International Airport [29]. This system is built on biometric verification and blockchain. It allows passenger to travel through passport control gate without manual check with holding up in lines. By deploying blockchain-based border control system, the biometric information of fingerprints, iris scans, facial recognition can be combined with the data currently stored in chips on existing e-passports to generate a digital passport. Blockchain technology ensures that the information stored on digital passport can only be accessed by own passenger and anyone else he specifically grants. Moreover, while the private digital data has become the property of third parties over last decades, the Dubai Airport's blockchain-based system will not only make international passenger quicker and safer, but it also gives people back control of their personal data. By using the self-sovereign identity based on blockchain technology, the data in digital passport are ensured to be only controlled by its owner, which enables the privacy of the individual kept. The usage of blockchain at Dubai Airport could give the benchmark for convincing leverage of blockchain innovation.

*B. Big Data and Artificial Intelligence*

According to study by research group IDC [30], in 2016, the online world produced 16 zettabytes (ZB) of data, but only 1% was analyzed due to lack of trust and transparency in how data is used. Seeing that there's a lot of data which sits dormant. They also predict that the world's data hits 163ZB a year by 2025, which requires safeguarding for being used for truly vital goals, at least on some level. Whereas data itself is a huge benefit industry, machine learning (ML) and Artificial Intelligence (AI) are also emerging industries that are set to impact every part of the society and economy in the coming years. ML and AI are robust technologies providing an opportunity to dramatically accelerate the capabilities of computing systems. Rather than being constrained by the parameters of its programming, these techniques can dynamically learn, find patterns, adapt, predict, and evolve. Additionally, AI needs data to be accurate and diverse. AI models have currently limited accuracy and usability without appropriate data. Owing to most of the world's giant data is used, AI is being held back from its power and potential. Unfortunately, only a handful of companies today have both Data and AI, like Facebook and Google, since sharing data amongst parties currently suffers from a bundle of issues, such as centralized storage, no control over data usage and lack of audit trial for compliance once data supplied by providers, lack of framework for trust, consent, adequately secured and regulation.

The climb of AI and ML has therefore been strongly correlated with the exponential upward trend in computing power afforded by cloud computing and Big Data that has become available over the last decade. ML and other AI technologies can also have an outstanding effect on how businesses manage data. Large US and Chinese tech giants such as Google, Baidu (search); Amazon, JD.com, Alibaba (ecommerce); Facebook, Tencent (communication, social network, games) entered the AI world. They have created large centralized cloud computing data centers, pushed into AI and ML research, and stored vast volume of user data which stemmed from the extensive growth in people using connected apps and systems that record user actions and sensory data continuously (IoT). Most AI and ML techniques make use of artificial neural networks (ANNs), called learning systems, which mimic the basic functioning of the human brain. Since ANNs require large amount of data for training purposes, tech-savvy players in many industries have allotted substantial resources to generate powerful data centers and acquire large datasets.

Naturally, data security will remain a challenge for all industries. Therefore, enterprise and even government working with AI are dealing with majority of challenges due to structural issues such as widespread invasion of user privacy, isolated and un-sharable data, lack of dataset transparency, biased or manipulated AI programs, exploitable and alterable datasets. Specifically, taking Facebook as an indication, they are holding a giant data storage and have a strong profit motive to violate user privacy and use AI/ML to find patterns and relationships that can be directly monetized from user data. On the other hand, centralized datastores are substantial interests for hackers since they provide millions of user records per penetrating - they are considered as a single point of failure with a high reward for bad actors. Although the owners of large datastores frequently are incentivized to rise their data size, they are not still intended to share their own data once they recognize it as a competitive benefit. So, it is evident that datasets are lack of sharing and collaboration.

Since security and self-sovereign ownership, decentralization and transparency are the outstanding properties of blockchain, there is the potential for generating significant benefits from incorporating the two technologies together. Additionally, collected datasets are important for training AI/ML algorithms, they should be accurate, immutable and adequate. With blockchain-based recording data technique, data are not saved until they go through a consensus mechanism, which ensures that the data being recorded is confirmed and accurate. Consequently, the data recorded are not only time-stamped, un-hackable and cryptographically signed, but they also provide the sharable ability for private and public entities. This utilization of blockchain can make datasets enhanced complete and diverse that can reduce bias for AI/ML.

Synapse AI project [31], built on Ethereum, creates a decentralized AI network platform where data contributors have awareness of their data on the network and get a compensation for their contribution. A cyclical economy is the aim of the project, in which agents contribute their data, then this data is pooled and used to create models that agents can consume. This cyclical process allows agents in the world to exponentially increase their capabilities by compounding their knowledge of the world. The SYN tokens are used for payments in this platform, for bonding to ensure quality is maintained, and for staking to support services.

As stated in Effect Network's whitepaper, the Effect.AI [32] platform leverages the NEO blockchain and is fueled by a network NEP-5 token called EFX to provide a decentralized runtime to AI applications. The 3 phases of The Effect Network, namely Force, Smart Market and Power component, have a low barrier of entry and allows anyone in the world to perform a wide range of tasks and receive fair payment. It gives AI developers and businesses access to a large workforce of human intelligence to train AI/ML algorithms. Each algorithm has its own wallet to allow for easy acceptance of transactions. People can also offer and buy AI algorithms as a service through Effect Smart Market. Meanwhile, Effect Power component provides AI models with a decentralized and distributed computational platform that runs popular machine learning and deep learning frameworks like MXNet, TensorFlow, Caffe.

Comparatively, Nebula AI [33] is a decentralized blockchain integrated with AI and sharing economies. With this blockchain-based AI solution, developers can deploy their AI applications easily on a blockchain platform. The integrated API/SDK and payment services allow the developers to earn revenue based on the AI smart contract. Nebula AI also allows the miners to contribute their GPU hash power to AI computing power which allow for a variety of AI applications.

AI startups and researchers have amazing algorithms, but lack data. The opposite is true for many large corporations. Ocean Protocol [34] is the bridge between these two sides, aims to unlock datasets, incentivizes for a large supply of relevant AI data and services. It is a decentralized protocol and network of AI ecosystem, which is supported by a Singapore based non-profit foundation. Their mandate is to equalize the opportunity to access data, provide data governance, encourage the AI network ecosystem growth and take measures to ensure that the platform gets ever more trust, secure, decentralized with time. So that, a much wider range of AI practitioners can create value from it and in turn spread the power of data.

*C. Security & Networking*

Obviously, the core features of blockchain technology are tamper-resistance and distributed system. When it comes to the application that best applies to the distributed ledger, security traceability becomes one of the fields most demonstrating the features of blockchain, with the most recognized at 85.7%, according to [35].

The intrusion detection system (IDS) is implemented to help organizations recognize cyber-attacks and alert about potential intrusions, incidents in a timely manner. Such detection systems have shown their capabilities of identifying cyber threats and protecting the network by monitoring system and network events, behaviors for any suspicious sign of possible incidents. However, cyberattacks have been getting more complicated and sophisticated, a single IDS meets many limitations of collecting and analyzing the target objects, traffic network and record information to generate an accuracy intrusion alarm. This is due to the detector can have false negatives or false positives. It is also specialized for certain types of threats that leads lack of view of the whole attack, for example, some intrusions could have multiple symptoms. To improve the recognition feature of a single IDS, many studies proposed the architecture of collaborative IDS (CIDS) which enables IDS entities communicate with each other to exchange data for monitoring and identifying anomalies, such as EMERALD [36], COSSACK [37], DOMINO [38], Netbait [39], PIER [40].

Nevertheless, CIDS have two main challenges of trust management and data sharing. Because not all IDS nodes desire to share their information regarding data privacy concerns, it is also hard to evaluate the trustworthiness of an IDS entity in distributed environment [41]. Meanwhile blockchain has been gathering a lot of attention of researcher's owing to its prospects of multi-sector adaptability breaking through the finance domain where it initially started. With trust, secure, transparent and incentive characteristics, blockchain can provide reliable connectivity for data exchange among collaborating parties in CIDS. By taking advantage of blockchain properties, Alexopoulos et al. [42] introduced the intersection of CIDSs and blockchain where blockchain provides the way of enhancing trust among IDS nodes. They proposed the idea of generic architecture to secure the generated alarms by various CIDS nodes and guarantee only trust notification or alerts would be transferred to the others. On the one hand, this initial design can be addressed the main challenges of data exchange and trust management by consensus algorithm in collaborative environment. On the other hand, their proposal needs to be implemented and resolve some challenges concerning the consolidation of blockchain and IDS approach from some inherent limitations separately.

Concerning to the rising technologies, Software Defined Networking (SDN) is an outstanding developing area with some exciting features such as the capabilities of controlling over network infrastructure and decoupling of control and data plane. In spite of its flexibility in network management, SDN still faces with multiple current and upcoming security threats associated with its deployment, such as rogue element accepted by a SDN may be able to view, surveil, copy, alter, disrupt communications over the entire network. Securechain [43], built on the Ethereum blockchain, is the solution that introduces security gateway into SDN, whilst creating a forensically auditable and unchangeable log of events. Because of blockchain's characteristics, this solution keeps the SDN safe from attacks, and gives the potential to set up automatic, programmable rules on what is - and what is not - acceptable on the SDN at any time and by whom. Securechain operation supports two features, including intergrating a new device to the SDN, and rejecting rogue element using a defined Command Wallet storing authorised and trusted entities. In the case of adding a device into the SDN, only requests sent from whitelisted wallets are stored in the blockchain and then being polled by SDN to deal with. With a valid instruction, the instruction can be implemented and the valid network element added, then the SDN controller is informed to allow the device to start functioning and be part of the SDN. In the second scenario, rejection of a rogue element takes place when requests are sent from a wallet not in the allowed list, and/or without the correct code within to attempt to add a rogue device. As stemmed from a hacker, the system rejects this instruction, and an alert is sent to the network admin to warn them of the intrusion, providing key details of the wallet ID of the hacker and timestamp. In addition, the rogue request is captured permanently in the blockchain, which allows for a security analysis of forensically auditable and unchangeable log with tamper-resistant. No longer could attackers attempt to cover their tracks by changing the history record of events since they cannot be tampered with without changing the blockchain.

In particular, DDoS attacks intend to cripple and drain the specific target by sending myriads of junk requests to a website or system until the target gets overwhelmed with requests and then crashes. Daphne Tuncer et al. [44] proposed a blockchain-based architecture of DDoS mitigation solutions in SDN using smart contracts across multiple domains. As a distributed and primarily public storage, this collaborative approach enables Autonomous Systems (ASes) to advertise DDoS attacks across multiple domains by white or blacklisted IP addresses notification, then enforce rules on the ASes-side to reduce DDoS impacts without transferring control of the network to a third party.

In [45], a decentralized firewall system powered by a novel malware detection engine is introduced. The authors built a firewall based on blockchain, while deep belief neural network (DBN) is used in detection engine to classify Portable Execution (PE) files as malicious or benign through analyzing representation of PE file in grayscale image format. All nodes of the network will serve as the detection engine by getting a unique trained model from central server. Any PE file would be broadcasted to all of nodes with their own engines, then running the classification progress to produce numeric value which represents the probability of the file being malicious. Every result of nodes was stored in blockchain as a transaction, thus the results given out by the nodes are infeasible to be tampered with. It is helpful for the broadcasted node to make the final decision by traversing the chain and performing a weighted average of the probability values added to the chain by other nodes in the network, corresponding to the same file. Probability value of each node is given a relevant weight as its measure of the node's trustworthy in the network. Upon every transaction, the probability of a node is compared to the average one over all the nodes, then the trust level will be updated based on the deviation. The integration of blockchain technique with the heuristic detection engine would provide full security support over a scalable network since any unauthorized modifications would be automatically invalidated.

Besides, the notion of smart city has been getting more attraction in few years, but there is no widely accepted definition of a smart city. The overarching mission of a smart city is to optimize city functions and drive economic growth while improving quality of life for its citizens using innovative technology and data analysis. In more detail, modern technologies are used to connect, protect, and enhance the lives of citizens and public safety. IoT sensors, video cameras, social media, and other inputs act as a brain and nervous system, providing the city operator and city dwell with constant feedback so they can make informed decisions properly. It is significant to assure these communications secure and trust due to requirements of private and sensitive information. Blockchain, which embodies availability, integrity, and non-repudiation, can be used as a reliable source of trust, as it is infeasible to exploit to major kinds of cyberattacks. As such, to grasp the effects of emerging blockchain technology on the development of smart cities, many recent researches are investigating the intersection of smart ecosystem and blockchain, as well as how blockchain can empower smart system with more satisfied privacy and security features.

According to [46], authors proposed a theorical security framework that consolidate smart devices with blockchain technology to enable a secure communication platform in a smart city. This finding takes advantage of blockchain to build resilient features against many cyberthreats, such as scalability, high fault tolerance, reliability and faster and efficient operation.

Blockcloud [47] is a service-centric blockchain architecture for IoT environment, where anyone can publish or subscribe a service to Blockcloud Marketplace, a decentralized application on Blockcloud, without central party involvement. Authors designed the consensus mechanism and ledger structure to provide trust, security and economically incentive in a decentralized way. Moreover, they also introduced a paradigm of service pricing and distribution that ensures anti-price cheating and fairness of sellers and buyers. To be more precise, both service providers and users of the network can earn profits with the same opportunity. Service providers will be rewarded with tokens to form a positive when contributing to the network. In addition, public feedback about the quality of services and reputation of providers is maintained during service trading. It is useful for customers to evaluate and choose reusable service on Blockcloud Marketplace.

## IV. DISCUSSION OF FUTURE DIRECTION

Emerging in recent years, blockchain technology is getting closer to its breakout moment day by day. 2018 witnessed an unprecedented rise in the development, working, and maturity of blockchain adoption. Data integrity, the strong point of Blockchain technique, is the reason why its use extends also to other domains and applications. Therefore, it has the full potential of revolutionizing data privacy, trust, security, and the relationship with individual information on the Internet.

After conducting a systematic review of many studies on data privacy and security sectors, we already found some findings that studied the possibility of utilizing blockchain in the context of network and security, IoT, smart environment, Big Data and ML/AI, also Identity Management system. This type of research trend will impact significantly in the future and can possibly be even more intriguing than cryptocurrencies. However, when more blockchain-based solutions are applied with large number of users, the scalability issues such as performance and latency obligate to be overcome for pervasive use of Blockchain.

## V. CONCLUSION

The personal and sensitive data should not be kept by third-parties, where they are potentially susceptible to attacks and misuse purposes without relevant permission. This paper has made an overview on Blockchain as a relevant answer for these privacy concerns with the elimination of third-party and capabilities of users to aware of their data that is being collected and used, some applications are also shown as examples. Whereas researchers must address some limitations and legal challenges in blockchain adoption, it is also significant to decide whether blockchain is the appropriate solution for a scenario of target application to keep up with the technological innovation.

ACKNOWLEDGEMENT

This work is funded by University of Information Technology, VNU-HCM under grant number of D1-2018-11.